\documentclass[paper,prd,reprint,preprintnumbers,nofootinbib,notitlepage,showpacs,showkeys]{revtex4-1}
\usepackage{latexsym,graphicx,amssymb,amsmath,mathrsfs} 

\usepackage{color}

\DeclareSymbolFont{bbold}{U}{bbold}{m}{n}
\DeclareSymbolFontAlphabet{\mathbbold}{bbold}

\newcommand{\be}{\begin{equation}}      
\newcommand{\ee}{\end{equation}}      
\newcommand{\bea}{\begin{eqnarray}}      
\newcommand{\eea}{\end{eqnarray}}    
     
\newcommand{\rt}[1]{{}}

\newcommand{\I}{\,\textrm{I}\,}  
\newcommand{\II}{\,\textrm{II}\,}  
\newcommand{\III}{\,\textrm{III}\,}  
\newcommand{\IV}{\,\textrm{IV}\,}  

\newcommand{\UV}{\,\textrm{UV}\,}
\newcommand{\STr}{\,\textrm{STr}\,}
\newcommand{\gf}{\,\textrm{gf}\,}
\newcommand{\gh}{\,\textrm{gh}\,}

\makeatletter
\renewcommand\appendix{\par
\setcounter{section}{0}%
\setcounter{subsection}{0}%
\gdef\thesection{\appendixname\space\@Alph\c@section}}

\long\def\unmarkedfootnote#1{{\long\def\@makefntext##1{##1}\footnotetext{#1}}}
\makeatother

\begin{document} 

\title{Fixed point structure of the Abelian Higgs model} 
\author{G. Fej\H{o}s}
\email{fejos@rcnp.osaka-u.ac.jp}
\affiliation{Research Center for Nuclear Physics, Osaka University, Ibaraki, Osaka 567-0047, Japan}
\affiliation{Theoretical Research Division, Nishina Center, RIKEN, Wako 351-0198, Japan}
\author{T. Hatsuda}
\email{thatsuda@riken.jp}
\affiliation{Theoretical Research Division, Nishina Center, RIKEN, Wako 351-0198, Japan}
\affiliation{iTHES Research Group, RIKEN, Wako 351-0198, Japan}

\begin{abstract}
{The order of the superconducting phase transition is analyzed via the functional renormalization group approach. For the first time, we derive fully analytic expressions for the $\beta$ functions of the charge and the self-coupling in the Abelian Higgs model with one complex scalar field in $d=3$ dimensions that support the existence of two charged fixed points: an infrared (IR) stable fixed point describing a second-order phase transition and a tritical fixed point controlling the region of the parameter space that is attracted by the former one. It is found that the region separating first- and second-order transitions can be uniquely characterized by the Ginzburg-Landau parameter $\kappa$, and the system undergoes a second order transition, only if $\kappa>\kappa_c \approx 0.62/\sqrt2$.}
\end{abstract}

\maketitle

\preprint{RIKEN-QHP-221}

\section{Introduction}

The order of the phase transition in the Abelian Higgs model with one complex scalar field became of interest because of the analyses of the spontaneous symmetry breaking due to radiative corrections in 3+1 dimensions \cite{coleman73}, and of a superconductor near the critical point with the dimensionally reduced Ginzburg-Landau theory \cite{halperin74}. It was suggested that fluctuations of the gauge field were of great importance and may even turn the second-order transition to first-order at least for strongly type-I superconductors. However, due to the fact that the temperature interval in conventional superconductors in which such fluctuations become important (i.e. the Ginzburg interval \cite{ginzburg60}) is of the order of $10^{-9}$ K,
the nature of the transition could not be resolved experimentally. We note that, in high-$T_{\rm c}$ superconductors, the Ginzburg interval can be of ${\cal O}(1)$ K; thus one can analyze the critical region more closely (see e.g. \cite{schneider05}).

Later on, it was argued, on the basis of a dual field theory \cite{kleinert82}, that the transition would keep the second-order nature if the Ginzburg-Landau parameter $\kappa=\lambda_L/\xi_c$ (the ratio of the London penetration depth and the correlation length) satisfied $\kappa>\kappa_c\approx 0.798/\sqrt{2}$. This has been confirmed by Monte Carlo (MC) simulations of the lattice model \cite{kajantie98}, with the critical value $\kappa_c=(0.76 \pm 0.04)/\sqrt{2}$ \cite{mo02}. Duality arguments indicate that the system belongs to the same universality class as of the three-dimensional XY-model \cite{kiometzis95,kleinert06,olsson98}.

The situation, however, could not be understood properly from renormalization group (RG) analyses and the fixed point structure of the theory. In perturbation theory with e.g. the minimal subtraction (MS) scheme, the divergence structure, which is necessary to be analyzed for the running couplings, can be calculated only in $d=4$. The corresponding $\epsilon$ expansion of the $\beta$ functions, extrapolated to $d=3$, does not produce an 
IR stable fixed point. By extending the order parameter to an $N$-component scalar field, one obtains $N\geq 183$ \cite{zinnjustin02} as a necessary condition of having charged fixed point(s) 
that may describe the second-order phase transition. Much effort has been made in establishing suitable RG approximation schemes both analytically and numerically \cite{kolnberger90,reuter94,bergerhoff96,bergerhoff96b,folk96,herbut96,freire01,kleinert03} to describe the phase structure: even though there have been indications of the existence of charged fixed points, it is believed that RG based analyses have not reached a satisfactory level of understanding the system, and certainly failed to produce a critical value of the Ginzburg-Landau parameter that is in accordance with MC simulations.

Here we wish to contribute in describing the superconducting phase transition from a RG point of view and present a completely analytical treatment that is capable of giving account of the corresponding second-order transition in the Abelian Higgs model. The method on the one hand reproduces the results of the $\epsilon$ expansion for the $\beta$ functions, and on the other hand for the first time provides fully analytic expressions directly in three dimensions that predict the existence of charged fixed points with a decent agreement for $\kappa_c$ given by MC simulations.

\section{Renormalization group flows}

We employ the functional renormalization group (FRG) formalism of quantum field theory. The method generalizes the idea of the Wilsonian RG in the sense that it provides not only the flow of individual coupling constants, but also the effective action itself. At the core of the FRG formalism lies the Wetterich equation \cite{wetterich93}:
\bea
\label{Eq:Wet}
\partial_k \Gamma_k = \frac12 \tilde{\partial}_k\STr \log (\Gamma_k^{(2)}+{\cal R}_k),
\eea
where ${\cal R}_k$ is a suitable IR regulator function (matrix) that suppresses modes with momenta $q\lesssim k$, leading $\Gamma_k$ to be the quantum effective action containing fluctuations with momenta $q \gtrsim k$. The operator $\tilde{\partial}_k$ acts only on ${\cal R}_k$, $\Gamma_k^{(2)}$ is the second functional derivative of $\Gamma_k$, and the supertrace (STr) operation has to be taken both in the functional and matrix sense. The scale parameter $k$ runs from a UV cutoff $k_{\UV}=\Lambda$ where no fluctuations are included and thus $\Gamma_{k\rightarrow \Lambda}\approx S$ ($S$ being the classical action), to $k=0$, reaching the quantum effective action itself, $\Gamma_{k=0}\equiv \Gamma$. For more details, the reader is referred to \cite{kopietz10}.

The Lagrangian of the $d$-dimensional Abelian Higgs model with one complex scalar field in Euclidean space reads as
\bea
\label{Eq:Lag}
{\cal L}&=&\frac12 A_i(-\partial^2\delta_{ij}+\partial_i \partial_j) A_j \nonumber\\
&+&(D_i\Phi)^\dagger D_i\Phi+m^2\Phi^\dagger \Phi+\frac{\lambda}{6}(\Phi^\dagger \Phi)^2,
\eea
where $D_i=\partial_i - ieA_i$ ($i=1,...d$) and $A_i$ is the $U(1)$ gauge field with $\Phi\equiv (\sigma+i\pi)/\sqrt2$ being the complex scalar. 
For gauge fixing, we employ the $R_\xi$ gauges \cite{fujikawa72}, and add ${\cal L}_{\gf}=\frac{1}{2\xi}(\partial_i A_i+\xi e\tilde{\sigma} \pi)^2$ to (\ref{Eq:Lag}), where $\tilde{\sigma}$ is a freely adjustable field, with $\xi$ being the 
 gauge fixing parameter.  This choice leads to the ghost dynamics of 
 ${\cal L}_{\gh}=c^*\left(-\partial^2+\xi e^2 \tilde{\sigma}\sigma\right)c$, and it considerably simplifies calculations if $\tilde{\sigma}$ is set to the classical field $\bar{\sigma}$ where one is interested in calculating the effective potential (not necessarily at the minimum), as it cancels for constant $\bar{\sigma}$ backgrounds the mixing between $A_i-\pi$ that originates from the covariant derivative.

Equation ({\ref{Eq:Wet}) manifests itself as an exact functional integro-differential equation, and after we let $\tilde{\partial}_k$ act on the log function, it can be represented diagrammatically as shown in Fig. 1. In this paper we employ the derivative expansion up to  ${\cal O}(\partial^2)$ to solve (\ref{Eq:Wet}) approximately, by taking into account the wave function renormalization of the fields via the formal rescalings $\Phi \rightarrow Z_{\Phi,k}^{1/2}\Phi$, $A_i \rightarrow Z_{A,k}^{1/2} A_i$, and also by performing multiplicative renormalization of the charge by $e \rightarrow Z_{e,k}/(Z_{A,k}^{1/2}Z_{\Phi,k})e$. It is well known in perturbation theory that $Z_{e,k}=Z_{\Phi,k}\equiv Z_k$; \footnote{This can be indeed verified by projecting (\ref{Eq:Wet}) to either the operator $\partial_i\Phi^\dagger \partial_i \Phi$ or $e^2A_iA_i \Phi^\dagger \Phi$, and finding the same coefficients.} thus we identify the scale-dependent charge $e_k = e/Z_{A,k}^{1/2}$.}
  
\begin{figure}[t]
\includegraphics[bb = 920 250 0 345,scale=0.4,angle=0]{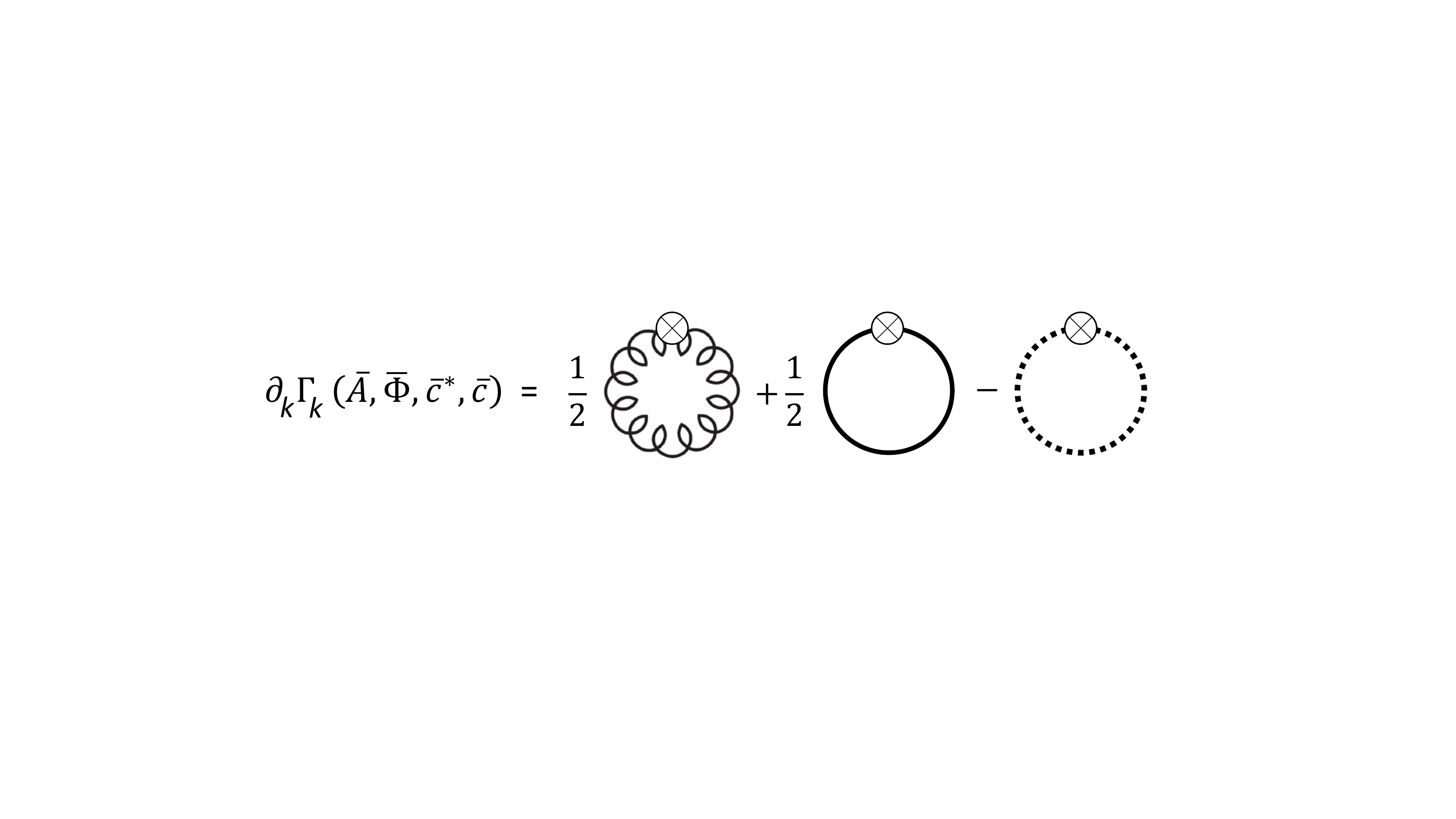}
\caption{Diagrammatic representation of the exact flow equation with appropriate kinematial factors for the Abelian Higgs model. The gauge, scalar, and ghost full propagators with the regulator functions ${\cal R}_k$ are shown by the coiled, solid and dashed lines, respectively.  Crosses represent insertions of $\partial_k {\cal R}_k$.}
\end{figure}  
  
Choosing $\tilde{\sigma}=\bar{\sigma}$, our ansatz for the $k$-dependent effective action is 
$\Gamma_k[\bar{A}_i,\bar{\sigma},\bar{\pi},\bar{c}^*,\bar{c}]=\int_x {\cal L}_k$, where
\bea
\label{Eq:Lk}
{\cal L}_k&=&\frac{Z_{A,k}}{2}\bar{A}_i\left(-\partial^2 \delta_{ij}+\partial_i \partial_j 
(1-\xi_k^{-1})\right)\bar{A}_j + \nonumber\\
&+&\frac{Z_k}{2}\bar{\sigma} (-\partial^2) \bar{\sigma} +\frac{Z_k}{2}\bar{\pi} (-\partial^2) \bar{\pi} + U_k(\bar{\sigma},\bar{\pi})\nonumber\\
&+&\bar{c}^*\left(-\partial^2+Z_k e_k^2\bar{\sigma}^2\right)\bar{c}-2Z_kZ_{A,k}^{1/2} e_k \bar{A_i} \bar{\pi} \partial_i \bar{\sigma} \nonumber\\
&+&\frac{Z_kZ_{A,k}}{2} e_k^2 \bar{A_i}\bar{A_i} \left(\bar{\sigma}^2
+\bar{\pi}^2 \right)+ \frac{\xi_k}{2}Z_k^2 e_k^2\bar{\sigma}^2 \bar{\pi}^2.
\eea
The $U_k(\bar{\sigma},\bar{\pi})$ function is a fully nonperturbative potential whose functional form should be determined by solving the respective flow equation. In the present paper, however, in order to determine the location of the possible tricritical point separating the first- and second-order phase transitions, 
we approximate $U_k(\bar{\sigma},\bar{\pi})$ near the critical point as
\bea
\label{Eq:Ukform}
U_k(\bar{\sigma},\bar{\pi})=\frac{Z_k m_k^2}{2}(\bar{\sigma}^2+\bar{\pi}^2)+\frac{Z_k^2\lambda_k}{24}(\bar{\sigma}^2+\bar{\pi}^2)^2.
\eea
This assumption with the ansatz (\ref{Eq:Lk}) reduces the Wetterich equation (\ref{Eq:Wet}) to coupled ordinary differential equations for the wave function renormalizations and couplings. In the following, the $\beta$ functions dictating the running couplings will be calculated up to {\cal $O$}$(e_k^4,e_k^2\lambda_k,\lambda_k^2$). Note that the wave function renormalization of the ghost field $Z_{c,k}$ contributes to the $\beta$ functions only in higher orders, and thus we set $Z_{c,k}=1$ in (\ref{Eq:Lk}). Also note that the gauge fixing parameter $\xi_k$ in FRG can be $k$ dependent due to the presence of the regulator \cite{gies12}.
 
For concrete calculations, we need to choose an ${\cal {R}}_k$ regulator matrix corresponding to the coupled dynamics of $\{A_i,\sigma,\pi,c^*,c\}$. Instead of defining it in advance, we first diagonalize the propagator matrix in a given background, and then associate a Litim type regulator \cite{litim01} for each eigenmode in the form of
\bea
\label{Eq:regulator}
\!\!{\cal R}_k^{(i)}(q)=Z_k^{(i)}R_k(q)\equiv Z_k^{(i)}(k^2-q^2)\Theta(k^2-q^2),
\label{Eq:Rk}
\eea
where $i=1,..., (d+4)$. The wave function renormalization (coefficient of $q^2$) of the $i$th eigenmode is denoted by $Z_k^{(i)}$, and it is a function of $Z_k$, $Z_{A,k}$ and $\xi_k$.

By plugging (\ref{Eq:Lk}) with $(\ref{Eq:Ukform})$ into (\ref{Eq:Wet}) and using the regulator (\ref{Eq:regulator}), one identifies the corresponding operators in the right-hand side and thus obtains the flow equations for wave function renormalizations and coupling constants. It is convenient to perform these identifications separately. First we evaluate (\ref{Eq:Wet}) for $\bar{\sigma}=$ const. with $\tilde{\sigma}=\bar{\sigma}$, which leads to the identification of the flow of $U_k$. We consider the system close to the critical temperature; thus $m_k^2 \approx 0$. Also, we neglect the effect of $\tilde{\partial}_k$ acting on $Z_k^{(i)}$ in the regulator (\ref{Eq:regulator}); this is justified in the present order of the approximation for the $\beta$ functions. Then one gets the following results in $d=4$ and $d=3$ dimensions:
\begin{subequations}
\bea
\label{Eq:lk4}
k\partial_k (Z_k^2\lambda_k)|_{d=4}&=&\frac{54 e_k^4 + 6e_k^2\lambda_k \xi_k 
+ 5\lambda_k^2}{24\pi^2} Z_k^2, \\
k\partial_k (Z_k^2\lambda_k)|_{d=3}&=&\frac{72 e_k^4 + 12e_k^2\lambda_k\xi_k
 +10 \lambda_k^2}{9\pi^2 k} Z_k^2.
\eea
\end{subequations}
Now, by taking a slowly varying field $\partial_i \bar{\sigma} \neq 0$ around $\bar{\sigma} \approx 0$, keeping the rest of the fields zero, the following results for the wave function renormalization $Z_k$ are obtained:
\begin{subequations}
\bea \label{Eq:Zk4}
k\partial_k Z_k|_{d=4}&=&\frac{3+\xi_k}{8\pi^2}e_k^2Z_k, \\
k\partial_k Z_k|_{d=3}&=&\frac{4(2+\xi_k)}{3\pi^2k}e_k^2Z_k.
\eea
\end{subequations}
Note that (\ref{Eq:lk4}) and (\ref{Eq:Zk4}) are the well-known 1-loop perturbative results for $d=4$. In order to obtain the $\beta$ functions and map the fixed point structure, we still have to identify the flow of $Z_{A,k}$. It can be done by evaluating the flow equation (\ref{Eq:Wet}) up to quadratic terms for a field configuration where the only nonzero background fields are $\bar{A}_i(x)$.

\begin{figure}
\includegraphics[bb = 480 475 120 550,scale=0.75]{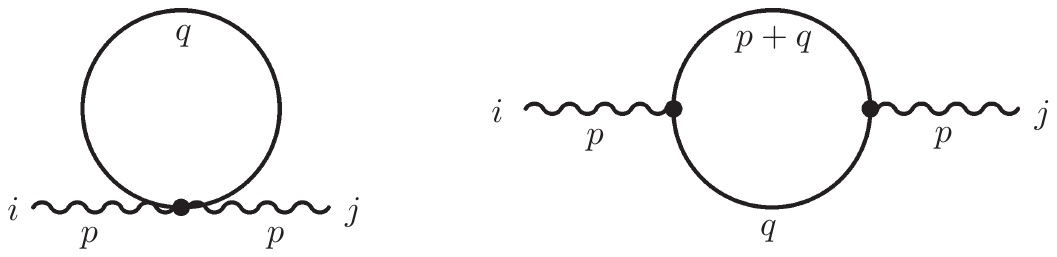}
\caption{1-loop diagrams that contribute to the 2-point gauge vertex.}
\end{figure}

The main issue of the FRG formalism at this step is the explicit breaking of local gauge invariance due to the momentum cutoff in the regulator (\ref{Eq:Rk}) for nonzero $k$. There is an extended literature on how the situation can be handled \cite{litim98,gies12}: the standard procedure consists of taking into account regulator modified Ward-Takahashi identities (mWTI) and first solving the flows of independent operators that are not constrained by them. As a second step, one lets the dependent operators such as the photon mass and longitudinal photon components flow by the mWTI rather than their respective flow equations. This is to ensure gauge invariance when all fluctuations are integrated out ($k \rightarrow 0$). In the present paper, we use a more heuristic approach, in which (i) we drop all contributions in the flow equation that is meant to produce a photon mass term since it is clearly an artifact due to the momentum cutoff \cite{herbut07}, and (ii) we determine the optimal gauge fixing parameter $\xi_k$ at the level of the approximation, by requiring  consistency of the flow equation and the ansatz (\ref{Eq:Lk}).

If we are to seek for the coefficient of the term quadratic in $\bar{A}$ to obtain the flow equation of $Z_{A,k}$, based on the 1-loop structure of (\ref{Eq:Wet}), two diagrams in Fig. 2 have to be taken into account. This results in
\bea
\!\!\!\!\partial_k \Gamma_k|_{\bar{A}}&=&\frac{e_k^2Z_{A,k}}{2} \int_p \bar{A}_i(-p) \bar{A}_j(p)\times \tilde{\partial}_k \Bigg[\int_q \frac{2\delta_{ij}}{q^2+R_k(q)} \nonumber\\
&-&\int_q \frac{(p+2q)_i (p+2q)_j}{(q^2+R_k(q))((p+q)^2+R_k(p+q))}\Bigg],
\eea
which, after discarding the photon mass term as mentioned in (i), simplifies to
\bea
\!\!\!\!\partial_k \Gamma_k|_{\bar{A}}&=&\frac12 \Big(-\frac{8e_k^2 Z_{A,k} k^{d-5}\Omega_d}{d(d+2)}\Big) \int_p \bar{A}_i(-p)\bar{A}_j(p)\times \nonumber\\
&&\Big(p^2\delta_{ij}-\frac{d-2}{2}p_i p_j\Big)+{\cal O}(p^4),
\eea
where $\Omega_d^{-1}=2^{d-1}\pi^{d/2}\Gamma(d/2)$. In the derivative expansion up to ${\cal O}(\partial^2)$, this leads to
\begin{subequations}
\label{Eq:ZAk}
\bea
\label{Eq:ZAk4}
\!\!\!\!\!\!\partial_k \Gamma_k|_{\bar{A},d=4}&=&\frac12 
\left(-\frac{e_k^2Z_{A,k}}{24\pi^2k}\right)\!\! \int_x \bar{A}_i(-\partial^2\delta_{ij}
+\partial_i\partial_j) \bar{A}_j,
\\
\label{Eq:ZAk3}
\!\!\!\!\!\!\partial_k \Gamma_k|_{\bar{A},d=3}&=&\frac12 
\left(-\frac{4e_k^2Z_{A,k}}{15\pi^2k^2}\right)\!\! \int_x \bar{A}_i (-\partial^2\delta_{ij}
+\frac12\partial_i\partial_j) \bar{A}_j. \nonumber\\
\eea
\end{subequations}
Note that (\ref{Eq:ZAk4}) is consistent with (\ref{Eq:Lk}) 
only if $\xi_k = \xi_{k=\Lambda} {Z_{A,k}}$.
This actually means that we do not let the longitudinal part flow with respect to $k$, in accordance with perturbation theory. A special choice is the Landau gauge $\xi_{k}=0$, which solely carries no $k$ dependence and thus turns out to be a fixed point of the RG flow, as it is known from earlier works \cite{litim98,gies12}.  
However, the choice of $\xi_{k=\Lambda}$ is not of any importance for $d=4$, since (\ref{Eq:ZAk4}) shows that 
$k\partial_k Z_{A,k}|_{d=4}=-e_k^2Z_{A,k}/24\pi^2$, which is the well-known result of perturbation theory, and thus the flows of $\lambda_k$ and $e_k^2$ are $\xi_{k}$ independent:
\begin{subequations}
\label{Eq:beta4}
\bea
k\partial_k \lambda_k|_{d=4}&=&\frac{54e_k^4-18e_k^2\lambda_k+5\lambda_k^2}{24\pi^2}, \\
k\partial_k e_k^2|_{d=4}&=&\frac{e_k^4}{24\pi^2}.
\eea
\end{subequations}
Also note that (\ref{Eq:beta4}) leads precisely to the standard results of the $\epsilon$ expansion: in $d=4-\epsilon$ dimensions, runnings of the dimensionless couplings, $\bar{\lambda}_k=\lambda_k/k^\epsilon$ and $\bar{e}_k^2=e_k^2/k^\epsilon$, are dictated by the respective $\beta_\lambda$ and $\beta_{e^2}$ functions:
\begin{subequations}
\label{Eq:beta4b}
\bea
\!\!\!\!\!\!\!\!\!\!\!\!\beta_\lambda|_{d=4-\epsilon}&=& k\partial_k \bar{\lambda}_k=-\epsilon \bar{\lambda}_k + \frac{54\bar{e}_k^4-18\bar{e}_k^2\bar{\lambda}_k+5\bar{\lambda}_k^2}{24\pi^2}, \\
\!\!\!\!\!\!\!\!\!\!\!\!\beta_{e^2}|_{d=4-\epsilon}&=&k\partial_k \bar{e}_k^2=-\epsilon \bar{e}_k^2+\frac{\bar{e}_k^4}{24\pi^2}.
\eea
\end{subequations}
As mentioned in the introduction, (\ref{Eq:beta4b}) shows no sign of an infrared stable fixed point extrapolated to $\epsilon=1$, and (incorrectly) signals a first-order transition in the whole range of the parameter space.

\begin{figure}
\includegraphics[bb = 50 80 590 810,scale=0.35,angle=-90]{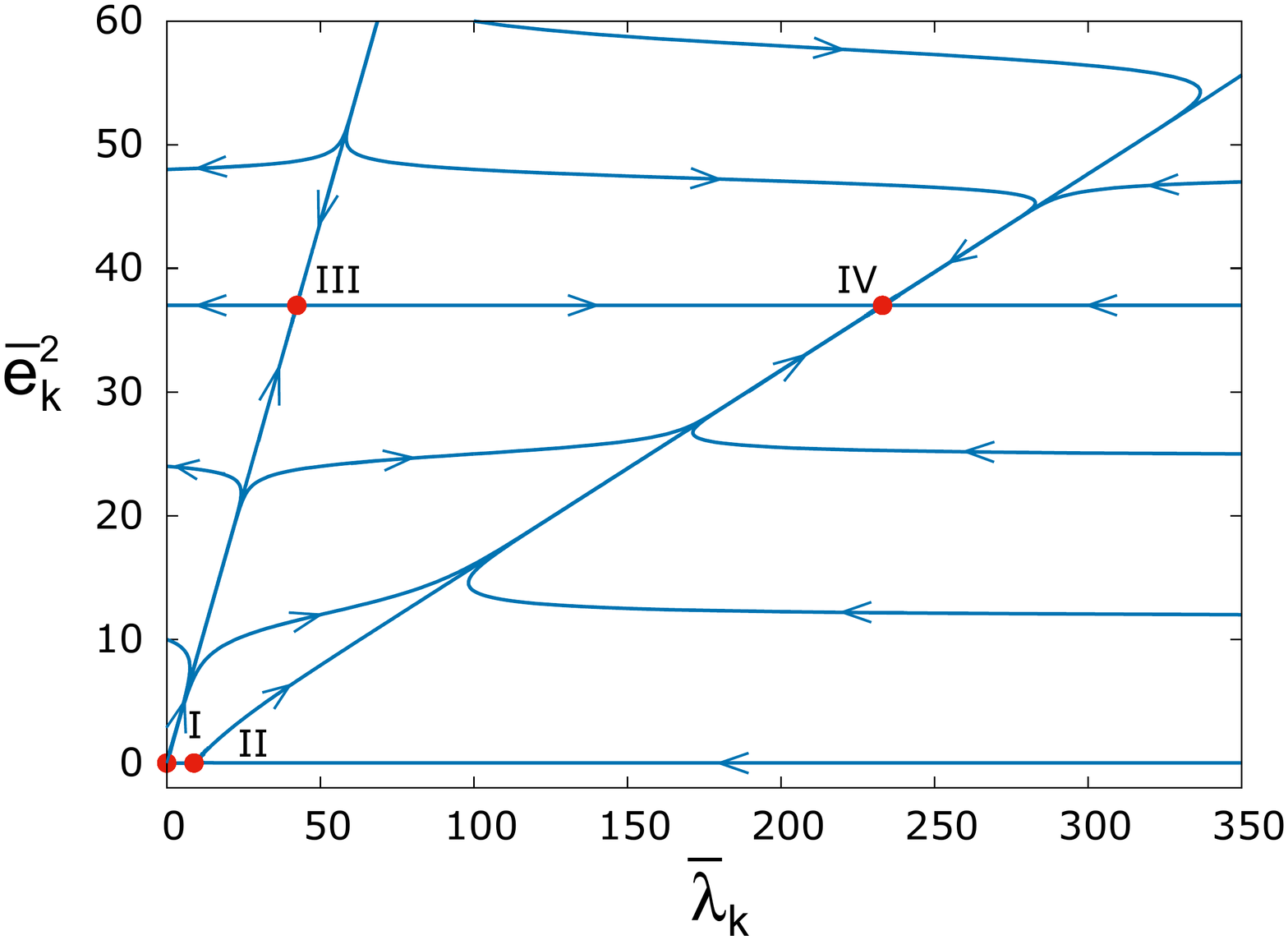}
\caption{Fixed points in the Abelian Higgs model for $d=3$. I and II are the Gaussian and Wilson-Fisher fixed points, respectively. III and IV are charged fixed points describing tricriticality and the phase transition itself.}
\end{figure}

Going back to (\ref{Eq:ZAk3}), the most important observation is that within the present approximation it is compatible with (\ref{Eq:Lk}), if and only if $\xi_k=2$ is satisfied as a fixed point. This uniquely determines $k\partial_k Z_{A,k}|_{d=3}=-4e_k^2Z_{A,k}/15\pi^2k$;
therefore, the $\beta$ functions directly in $d=3$, without using the $\epsilon$ expansion, read
\begin{subequations}
\label{Eq:beta3}
\bea
\beta_{\lambda}|_{d=3}&=&-\bar{\lambda}_k+\frac{72\bar{e}_k^4-72 \bar{e}_k^2\bar{\lambda}_k+10\bar{\lambda}_k^2}{9\pi^2}, \\
\beta_{e^2}|_{d=3}&=&-\bar{e}_k^2+\frac{4}{15\pi^2}\bar{e}_k^4.
\eea
\end{subequations}

\begin{figure}
\includegraphics[bb = 920 235 0 355,scale=0.42,angle=0]{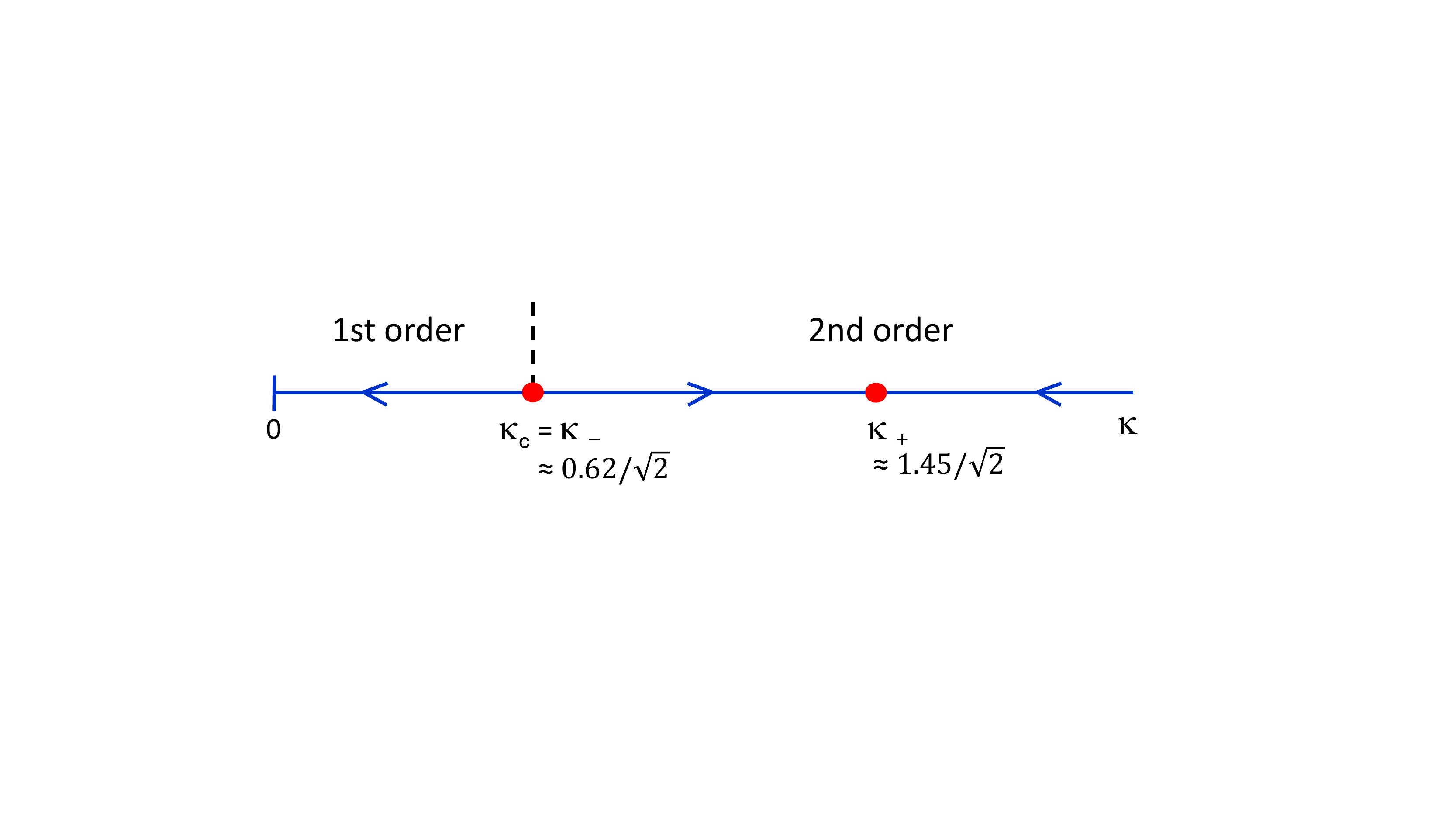}
\caption{Flow of the Ginzburg-Landau parameter. If $\kappa>\kappa_-$, the phase transition is of second order.}
\end{figure}

\section{Fixed points in d=3}

Unlike the results of the $\epsilon$ expansion (\ref{Eq:beta4b}), $\beta$ functions (\ref{Eq:beta3}) do show a charged infrared stable fixed point. Setting $\beta_{\lambda}|_{d=3}=0$, $\beta_{e^2}|_{d=3}=0$, one analytically solves the coupled equations and obtains the following fixed points:
\begin{subequations}
\bea
\bar{\lambda}_{\I}&=&0, \qquad \qquad \qquad \qquad \quad \bar{e}_{\I}^2=0, \\
\bar{\lambda}_{\II}&=&\frac{9\pi^2}{10}, \qquad \qquad \qquad \qquad \!\!\!\!\bar{e}_{\II}^2=0, \\
\bar{\lambda}_{\III}&=&\frac{9\pi^2}{20}(31-\sqrt{461}), \qquad \!\! \bar{e}_{\III}^2=\frac{15\pi^2}{4}, \\
\bar{\lambda}_{\IV}&=&\frac{9\pi^2}{20}(31+\sqrt{461}), \qquad \!\! \bar{e}_{\IV}^2=\frac{15\pi^2}{4}.
\eea
\end{subequations}
Fixed points I and II are the Gaussian and Wilson-Fisher fixed points, respectively; III in turn is a so-called tricritical fixed point, having one unstable direction, that is responsible for the separation between first- and second-order transitions, by controlling the region of the parameter space that is attracted by fixed point IV (see Fig. 3). The latter one is IR stable in both directions, and thus corresponds to a second-order transition.

To determine the region of the parameter space that belongs to a second-order transition, we recall that the Ginzburg-Landau parameter at scale-$k$ in the present approximation is $\kappa_k=\sqrt{{\lambda_k}/{6e_k^2}}\equiv \sqrt{{\bar{\lambda}_k}/{6\bar{e}_k^2}}$. Using (\ref{Eq:beta3}), the flow of $\kappa_k^2$ turns out to be
\bea
\label{Eq:kappaflow}
k\partial_k \kappa_k^2=\frac{4\bar{e}_k^2}{15\pi^2}(25\kappa^4_k-31\kappa^2_k+5).
\eea
In the physical region (i.e. $\kappa_k>0$),  (\ref{Eq:kappaflow}) has two fixed points:
\vspace{-0.2cm}
\bea
\kappa_{\pm}=\sqrt{\frac{31\pm \sqrt{461}}{25}}/\sqrt2;
\eea
see also Fig. 4. One observes that if the Ginzburg-Landau parameter satisfies $\kappa>\kappa_-\equiv \kappa_c\approx 0.62/\sqrt2$, then it flows toward $\kappa_+\approx 1.45/\sqrt2$, which is the universal value of the ratio $\lambda_L/\xi_c$ at the transition point. However, if $\kappa<\kappa_c$, the couplings are not part of the attraction region of the IR stable fixed point. We thus achieved our goal and showed (analytically) that if $\kappa>\kappa_c$, the couplings flow to the IR stable fixed point and thus the superconducting transition is of second order. First-order transitions can only occur if $\kappa <\kappa_c$.

\vspace{-0.4cm}
\section{Conclusions}
\vspace{-0.2cm}
We have discussed the superconducting phase transition within the functional renormalization group approach. Flow equations for the Abelian Higgs model with one complex scalar field have been solved at ${\cal O} (\partial^2$) level of accuracy of the derivative expansion. Our method has reproduced the results of the $\epsilon$ expansion for the $\beta$ functions and also provided fully new analytic formulas in $d=3$. We have seen that matching the flow equation with the ansatz of the effective action selects appropriate choices of gauge fixing parameters within the approximation. The resultant gauge choice for $d=3$ and the associated $\beta$ functions of the charge and the self-coupling show both tricritial and infrared stable charged fixed points. The obtained critical value of the Ginzburg-Landau parameter ($\kappa_c\approx 0.62/\sqrt2$) that separates first- and second-order transitions is in fair agreement with Monte Carlo simulations. 

The resultant $\kappa_c$ could be made more accurate by solving the flow of the scalar potential nonperturbatively and taking into account the field dependence of the wave function renormalizations. Generalization to an $N$ component scalar field is straightforward with the result $\kappa_c=\sqrt{\frac{N+30-\sqrt{N^2-40N+500}}{5(N+4)}}/\sqrt2$. These represent future works and the details will be reported elsewhere.

\begin{widetext}

\section*{Acknowledgements}
G. F. was supported by the Foreign Postdoctoral Research Program of RIKEN. T. H. was partially supported by the RIKEN iTHES Research Project.

\end{widetext}

\makeatletter
\@addtoreset{equation}{section}
\makeatother

\end{document}